\documentclass[twocolumn]{elsart3p}

\usepackage{epsfig}

\usepackage{amssymb}

\begin{document}

\begin{frontmatter}



\title{Measurement of Giant Dipole Resonance width at low temperature:
A new experimental perspective}


\author[label1]{S. Mukhopadhyay},
\author[label1]{Deepak Pandit},
\author[label1]{Surajit Pal},
\author[label2]{Srijit Bhattacharya},
\author[label3]{A. De},
\author[label1]{S. Bhattacharya},
\author[label1]{C. Bhattacharya},
\author[label1]{K. Banerjee},
\author[label1]{S. Kundu},
\author[label1]{T. K. Rana},
\author[label1]{G. Mukherjee},
\author[label1]{R. Pandey},
\author[label1]{M. Gohil},
\author[label1]{H. Pai},
\author[label1]{J. K. Meena}
\author {and}
\author[label1]{S. R. Banerjee\corauthref{cor}}
\corauth[cor]{Corresponding author.}
\ead{srb@vecc.gov.in}


\address[label1]{Variable Energy Cyclotron Centre, 1/AF-Bidhannagar, Kolkata-700064, India}
\address[label2]{Department of Physics, Barasat Govt. College, Barasat, N 24 Pgs, Kolkata - 700124, India }
\address[label3]{Department of Physics, Raniganj Girls' College, Raniganj - 713358, India}

\begin{abstract}
The systematic evolution of the giant dipole resonance (GDR) width in the temperature region of  0.9 $\sim$ 1.4 MeV has been measured experimentally for $^{119}$Sb using alpha induced fusion reaction and employing the LAMBDA high energy photon spectrometer. The temperatures have been precisely determined by simultaneously extracting the vital level density parameter from the neutron evaporation spectrum and the angular momentum from gamma multiplicity filter using a realistic approach. The systematic trend of the data seems to disagree with the thermal shape fluctuation model (TSFM).
The model predicts the gradual increase of GDR width from its ground state value for  $T>$ 0 MeV whereas the measured GDR widths appear to remain constant at the ground state value till  $T\sim$  1 MeV  and increase thereafter indicating towards a failure of the adiabatic assumption of the model at low temperature.   

\end{abstract}

\end{frontmatter}


In recent years, great interests are being shown to understand the correct 
description of damping mechanisms contributing to the Giant Dipole Resonance 
(GDR) width at low temperatures ($T$) \cite{heck03, cam03, dipu1}. 
The damping of such giant collective vibration inside the nuclear medium occurs either due to 
escape of resonance energy by means of particle or photon emission (escape width) 
or due to its redistribution in other degrees of freedom within the system (spreading width) \cite{hara01}.
In medium and heavy nuclei, it turns out that the escape width only account for a small fraction 
and the major contribution of the large resonance width comes from the spreading width \cite{bort01,don01}.
The general trend of the resonance width, as deduced from the lorentzian fit to the cross-section data, 
has been found to be smallest for the closed shell nuclei and larger for the nuclei between shells \cite{hara01}. 
However, it needs to be mentioned that for deformed nuclei, the width was obtained by fitting 
one single Lorentzian. Such fits never resulted in a systematic mass dependence of the width. 
Recently, an empirical formula has been derived for the spreading width by separating the deformation 
induced widening from the spreading effect and requiring the integrated Lorentzian curves 
to fulfil the dipole sum rule \cite{jun08}. The relation has been found to hold good for the widths of the different GDR
components corresponding to the three axes of a deformed nucleus in general \cite{jun08, erh10}. 
The inclusion of the deformation in describing the apparent GDR width is also supported by the recent 
experimental \cite{sre05} and theoretical \cite{del10} development in the description of the nuclear ground states. 

The GDR, built on excited state, is an important experimental tool since it couples directly to the nuclear 
shape and the investigation of its strength distribution gives a direct access to the nuclear deformation. 
Owing to this property, it has been applied to study Jacobi shape transition \cite{Maj04, Dipu2} and hyper-deformation  \cite{Dipu2}
in alpha cluster nuclei. This triaxiality is caused by J-driven deformation in high J and T regime and 
could be measured experimentally since the shapes are characterized by very large deformation ($\beta \gtrsim$ 0.6).
The experimental results have been substantiated with a theoretical calculation based on the thermal shape fluctuation model [TSFM] \cite{alh88}, 
which takes into account the J-driven deformation and T-driven shape fluctuation.
For a triaxial non-rotating nucleus, the GDR strength function is a superposition of three Lorentzians 
that correspond to the vibration of the nucleus along each of the semi-axes \cite{bus91}. The resonance energy corresponding 
to the each axis is obtained using the Hill-Wheeler parametrization
$E_k=E_0\exp\left[-{(\sqrt{5/4\pi}\,)}\beta \cos{(\gamma-2\pi k/3)}\right]$
while the widths are calculated applying the power law
$\Gamma_k=\Gamma_0 (E_k/E_0)^{1.6}$ \cite{alh93}. E$_0$ and  $\Gamma_0$ are the parameters for a spherical nucleus with mass A.
Further, at very high angular frequencies, these three GDR components split (the ones perpendicular to the spin axis) due to 
Coriolis effect as the GDR vibrations in a nucleus couple with its rotation when viewed from a non-rotating frame giving rise to five GDR components altogether \cite{Nee82, Maj04, Dipu2}.
Finally, the GDR cross-section is calculated by taking into account the large amplitude thermal 
fluctuations using a Boltzmann probability e$^{-F(\beta, \gamma)/T}$ with the volume
element $\beta^4 \sin(3\gamma) d\beta d\gamma$, where $F$ is the free energy \cite{alh93}.
However, for small deformations, experimentally it is not possible to measure the shape of the nucleus since 
thermal shape fluctuation smears out the associated splitting of the strength function resulting in an 
overall broadening of the distribution. Thus, only the apparent GDR widths are measured from the 
experiment using a statistical model analysis and compared with the TSFM, which also provides the 
apparent width of the GDR (in turn the shape of the nucleus). A systematic study of the thermal fluctuation model revealed the existence 
of a universal scaling law for the apparent width of the GDR for all T, J and A \cite{kus98, sri08}.

The apparent width of the GDR, built on the excited states, has been found to increase 
monotonically ($\sim$$T$$^{1/2}$)  \cite{thoe04} beyond $T>$ 1.5 MeV. One should expect a gradual 
increase in the apparent GDR width from its ground state value ($T=$ 0 MeV) 
with the  increase in temperature as predicted by TSFM. 
However, the temperature region below 1.5 MeV has rarely been investigated to verify if such a behavior is really true.
In Sn and nearby nuclei ( $A\sim$ 120), mostly investigated so far, only a single apparent GDR width measurement 
exists for  $T<$ 1.2 MeV which lies well below the TSFM prediction \cite{heck03}.
On the other hand, the phonon damping model (PDM)\cite{dang03} which considers the coupling of the GDR phonon to particle-particle and hole-hole configurations as the mechanism for the increase of GDR width, without any need of $T$-driven shape fluctuations,  
attributes this suppression to thermal pairing which contributes even beyond 1 MeV. 
These two models clearly disagree with one another at temperatures below 1.5 MeV  
highlighting the importance of microscopic effects responsible for this unusual phenomenon. 
In order to address these issues and to test the validity of the theoretical models, 
a systematic comparison between experiment and theory over a range of 
temperature for several nuclei is required.

Experimentally, the measurement of GDR width at low temperature is 
very challenging due to the difficulties in achieving low 
excitation energy. Traditional heavy ion fusion reactions 
are limited to higher temperature due to the presence of 
Coulomb barrier in the entrance channel and are always 
associated with broad $J$ distributions. Inelastic scattering \cite{heck03, bau98, rama96}  
has been used as an alternative approach with the advantage 
that the angular momentum transfer will be relatively low,  
but, the excitation energy windows are uncertain to about at least 
10 MeV and hence, the estimated temperatures are less precise. 
Due to these reasons, very few and widely separated 
($\sim$ 0.25 MeV) data points with large error bars are available. 
In the present work, alpha induced fusion reactions 
with precise experimental techniques has been used to investigate the 
low temperature region. In these reactions, the description 
of excitation energies from where the GDR photons decay will 
be more precise and  the associated maximum angular momentum for  $A\sim$ 119 mass region will 
be rather small ($\sim$ 20$\hbar$).
	
In this Letter, we present the first systematic and precise experimental 
study of angular momentum gated apparent GDR width in the unexplored low temperature region
(0.9 - 1.4 MeV) for $^{119}$Sb using fusion reaction with alpha particles. 
The nuclear level density (NLD) parameter is a critical input for statistical model 
calculation and is also important for proper estimation of the
nuclear temperature \cite{hara01}. The NLD parameter has been measured experimentally from the 
neutron evaporation spectrum while the bremsstrahlung contribution 
has been estimated using the forward/backward $\gamma$-ray anisotropy \cite{kell99}. 
	
The experiment was performed at the Variable Energy 
Cyclotron Centre (VECC), Kolkata using  accelerated  alpha beams from the
K130 Cyclotron. A self-supporting 1 mg/cm$^2$ thick target 
of $^{115}$In (99$\%$ purity) was bombarded with beams of $^{4}$He. 
Three different beam energies of 30, 35 and 42 MeV were
used to form the compound nucleus $^{119}$Sb at the excitation 
energies of 31.4, 36.2 and 43.0 MeV, respectively. 
The LAMBDA high energy photon 
spectrometer \cite{supm} (98 large BaF$_2$ detectors 
arranged in two blocks of 7$\times$7 each) was used to measure the high 
energy gamma rays ($\geq$ 4 MeV) at the angles of 
55, 90 and 125 degrees with respect to the beam axis. The detector arrays were 
positioned at a distance of 50 cm from the target.
Since, the GDR parameters depend on both the excitation energy and the 
angular momentum populated, it is important to separate the two effects 
in order to understand their individual contribution.
Hence, along  with the 
LAMBDA spectrometer, a 50-element low energy $\gamma$ multiplicity filter \cite{dipu}  
was used (in coincidence with the high energy $\gamma$-rays) 
to estimate the angular momentum populated in the 
compound nucleus in an event-by-event mode as well as to get a fast
start trigger for 
time-of-flight (TOF) measurements. 
The filter was split into two blocks of 25 detectors each
which were placed on 
top and bottom of a specially designed scattering 
chamber at a distance of 5 cm from the target in a staggered 
castle type geometry. The TOF technique was used 
to discriminate the neutrons from the high energy $\gamma$-rays.
The pulse shape discrimination (PSD) technique 
was adopted to reject the pile-up events in the individual 
detector elements by measuring the charge deposition 
over two integrating time intervals  (30 ns and 2 $\mu$s) \cite{supm}. 
The neutron evaporation spectra were 
measured using seven liquid scintillator (BC501A, 5$''$ diameter and 7$''$ long)  based 
neutron time of flight detectors \cite{kb} in coincidence 
with the multiplicity filter. The neutron detectors were 
placed at the angles of
30, 45, 75, 90, 105, 120 and 150 degrees with 
respect to the beam direction and at a distance of 150 cm from the target.
The time resolution of the neutron detectors was typically about 1.2 ns 
which gives an energy resolution of about 15$\%$ at 1 MeV for the present setup.

\begin{figure}
\begin{center}
\includegraphics[height=9.0 cm, width=6.0 cm]{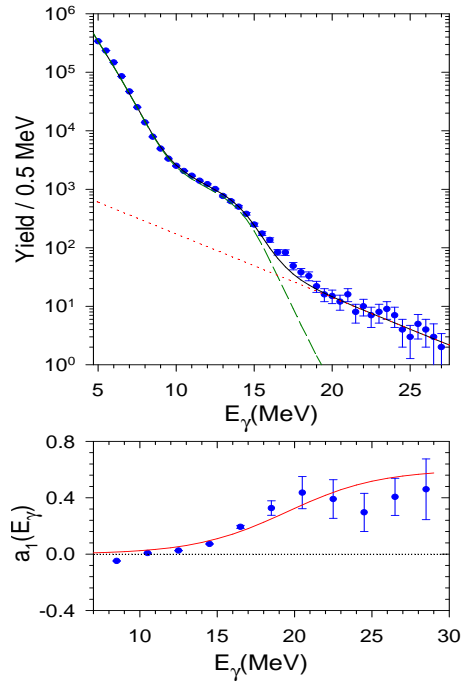}
\caption{\label{brem} [Color online] (top panel) The experimental gamma ray energy spectrum (blue circles) at 35 MeV is compared 
with the CASCADE prediction with bremsstrahlung component (black continuous line). The individual CASCADE (green dashed) 
and bremsstrahlung (red dotted) components are also shown. (bottom panel) The experimental a$_1$ coefficient (blue circles) is compared 
with the exponential fit (red continuous line) with the same E$_0$ parameter as used for the bremsstrahlung shape.}
\end{center}
\end{figure}

In alpha induced fusion reactions at these energies,  the populated 
angular momenta are quite low to have any effect on the widening of the GDR width.
The angular momentum dependent increase of the apparent GDR width starts showing up 
above a spin of 35$\hbar$ for  $A\sim$ 120 mass region ( $J\sim$ 0.6$A$$^{5/6}$) \cite{kus98}.
However, it is still important to 
measure the angular momentum to properly estimate the
temperature of the compound nucleus. The measured fold distribution from the 
multiplicity filter was mapped onto the angular momentum 
space using a Monte Carlo GEANT3 \cite{brun86} 
simulation. 
The procedure is described in detail in ref \cite{dipu}. 
The extracted angular momentum values for three different 
incident energies are listed in Table \ref{data}.

The energy of the evaporated neutrons has been measured 
using TOF technique whereas the neutron gamma 
discrimination was achieved by pulse shape discrimination 
(PSD) and TOF. The neutron TOF spectra were converted to 
neutron energy spectra using the prompt gamma peaks in the TOF spectra 
as the time reference. The efficiency correction for the neutron 
detectors were done using the Monte Carlo Computer code NEFF 
\cite{kb}. The evaporated neutron energy spectra, 
after transformation from laboratory frame to centre of 
mass frame, were compared with CASCADE \cite{cas} 
calculation using Chi-square minimization technique 
in the energy range of 2 - 8 MeV for the determination of the level density 
parameter ($\widetilde{a} = A/a$). The extracted inverse 
level density parameters ($a$) are listed in Table \ref{data} 
with an uncertainty of $\pm$ 0.4 MeV.

\begin{figure}
\begin{center}
\includegraphics[height=10.0 cm, width=7.0 cm]{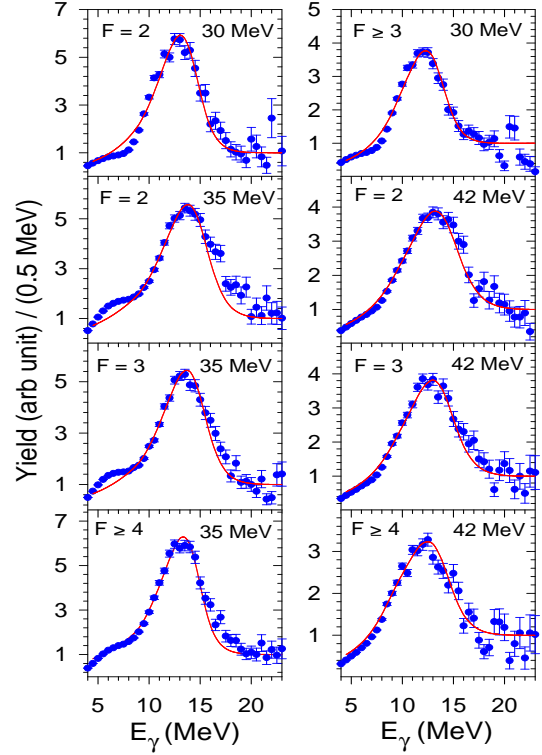}
\caption{\label{gdr} [Color online] 
Divided plots of the measured $\gamma$-spectra (blue circles) and the best fitted CASCADE calculations
(red continuous lines) for different folds (F) at incident energies of 30, 35 and 42 MeV. }
\end{center}
\end{figure}

The high energy $\gamma$-ray spectra were generated 
in the offline analysis using the cluster summing technique 
\cite{supm} in which each detector element was required 
to satisfy the prompt time gate and pulse shape 
discrimination gate. The measured high energy 
$\gamma$-ray spectra at 90$^{\circ}$ were compared with 
a modified version of the statistical model 
code CASCADE \cite{cas} along with a bremsstrahlung component.
The non statistical contributions to the experimental 
$\gamma$ spectra arising due to bremsstrahlung emission 
were parametrized using the relation 
$\sigma$$_{brem}$ = k/[C+exp(E$_{\gamma}$/E$_0$)] \cite{kell99}. 
The centre of mass $\gamma$-ray angular distributions were
assumed to have the form $\sigma(\theta)$=A$_{0}$[1 + a$_{1}$P$_{1}$(cos($\theta$)) + 
a$_{2}$P$_{2}$(cos($\theta$))] as the emission of gamma rays is dominated by electric dipole radiation \cite{kell99}. 
The a$_1$ coefficient should be zero for statistical emission, however, it is non zero 
for higher gamma energies due to bremsstrahlung emission.
The slope parameter (E$_0$) of the bremsstrahlung shape was extracted by simultaneously fitting the 
a$_1$ coefficient using the exponential function with the same slope parameter. 
The bremsstrahlung component as well as the a$_1$ 
coefficient for 35 MeV incident energy is shown in Fig \ref{brem}. 
The extracted value of the 
slope parameters are consistent with the systematics 
E$_0$ = 1.1[(E$_{Lab}$ - V$_{c}$)/A$_{p}$]$^{0.72}$, where E$_{Lab}$, 
V$_{c}$ and A$_{p}$ are the beam energy, coulomb barrier and 
the projectile mass respectively \cite{nif90}.
The Coulomb barrier in the studied reaction is 15.1 MeV. 
In the CASCADE calculation, the level density prescription 
of Ignatyuk \cite{igna75} has been taken with the asymptotic 
level density parameter as extracted from the  corresponding neutron 
evaporation spectrum. The simulated spin distributions deduced 
from the experimental multiplicity distributions were 
used as inputs for different folds. 
The predictions from the CASCADE calculations and 
the bremsstrahlung contributions were convoluted 
with the detector response and compared with the 
experimental gamma-ray spectra for different folds. 
The best fit was obtained using a $\chi$$^2$  
minimization technique in the region of 8 - 20 MeV.
In order to highlight the GDR region, both the data and calculated spectra were 
linearized by dividing with a statistical spectrum assuming constant E1 strength (Fig \ref{gdr}).
The average temperature was estimated using 
the relation $\left\langle T \right\rangle$=[(E$^*$ - E$_{rot}$ - 
E$_{GDR}$)/$\widetilde{a}$]$^{1/2}$  
where E$^*$ is the excitation energy. E$_{rot}$ is the
energy bound in the rotation at the average $J$ corresponding 
to a particular fold.  The GDR centroid energies (E$_{GDR}$) 
were found to be constant at around 15 MeV. 
The extracted parameters for different incident energies
are listed in Table \ref{data}.  
At this point, it needs to be mentioned that the nuclear deformation was not included in our analysis 
and we report on the extraction of the apparent GDR widths and compare them with the TSFM, 
which also provides the apparent width of the GDR including all shape fluctuations. 

\begin{figure}
\begin{center}
\vspace{0.3cm}
\includegraphics[height=4.5 cm, width=7.3 cm]{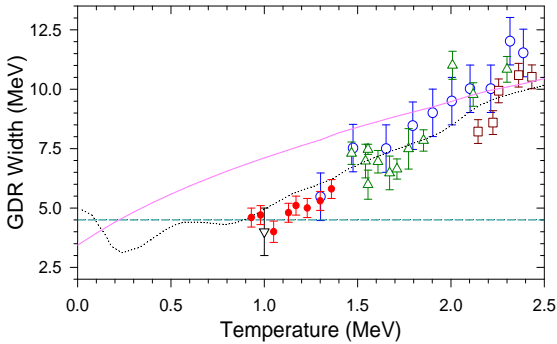}
\caption{\label{tsfm_1} [Color online] Plot of apparent GDR width with temperature. The red filled circles are the values for $^{119}$Sb deduced from the present work. 
For comparison, data for $^{120}$Sn from previous works are shown. The black downward triangle is from ref \cite{heck03}. 
The green upward triangles and blue open circles are from ref \cite{kus98}  while the brown open squares are from ref\cite{kell99}.
The pink continuous line represents the TSFM calculation \cite{kus98} while the black dotted line corresponds to the 
phonon damping model calculation \cite{dang03}. The green dashed line is the apparent ground state width of $^{119}$Sb (discussed in the text).}
\end{center}
\end{figure}


\begin{table*}
\caption{\label{data} GDR width and bremsstrahlung parameters at various folds for 30, 35 and 42 MeV  beam energies.} 
\vspace{0.25cm} 
\begin{center}		
\begin{tabular}{|c|c|c|c|c|c|c|c|}

\hline

E$_{Lab}$   & $\;$ Fold $\;$  &  $\left\langle J \right\rangle$ & E$_{rot}$& $\;$ $\left\langle T \right\rangle$ $\;$  & Apparent GDR  & E$_0$ & $a$ \\
$\;$MeV$\;$ &      &   $\hbar$ & $\;$ MeV $\;$  & $\;$ MeV  $\;$       &    width (MeV)  &  $\;$  MeV $\;$ & $\;$ MeV$^{-1}$  $\;$   \\ \hline
30     & 2    & $\;$ 13.5 $\pm$ 4.7 $\;$         & 2.5 &  0.98   & 4.7 $\pm$ 0.3 & 2.8 & 8.0    \\
    & $\geq$ 3    &	$\;$ 18.0 $\pm$ 5.2 $\;$ & 4.3 &  0.93   & 4.6 $\pm$ 0.3 & 2.8 &     8.0      \\ \hline
     & 2    &	$\;$ 12.2 $\pm$ 4.7 $\;$         & 2.1 &  1.17   & 5.1 $\pm$ 0.3 & 2.8 &   8.5     \\
35    & 3    &	$\;$ 15.1 $\pm$ 4.8 $\;$         & 3.1 &  1.13   & 4.8 $\pm$ 0.3 & 3.0 & 8.5      \\
       & $\geq$ 4    &	$\;$ 20.1 $\pm$ 5.2 $\;$ & 5.3 &  1.03   & 3.9 $\pm$ 0.3 & 3.0 & 7.9       \\ \hline
     & 2    &	$\;$ 16.2 $\pm$ 4.8 $\;$         & 3.5 &  1.36   & 5.8 $\pm$ 0.3 & 3.5 &   8.5        \\
42    & 3    &	$\;$ 19.1 $\pm$ 4.9 $\;$         & 4.8 &  1.30   & 5.3 $\pm$ 0.3 & 3.5 & 8.5        \\
       & $\geq$ 4   &	$\;$ 23.5 $\pm$ 5.3 $\;$   & 7.1 &  1 .23   & 5.0 $\pm$ 0.3 & 3.8 &8.5        \\ \hline

\hline
\end{tabular}
\end{center}		
\end{table*}

The  apparent GDR widths measured in the low temperature range of 0.9 - 1.4 MeV 
in the present study are shown in Fig \ref{tsfm_1} along with 
other measurements done earlier for $^{120}$Sn \cite{heck03, kus98, kell99}.
The continuous line represents the 
adiabatic thermal shape fluctuation calculation  \cite{kus98} 
at low spin. It is evident that the temperature dependence of the apparent GDR 
width determined from this experiment differs substantially from 
the adiabatic thermal shape calculation at low temperature. 
In  $A\sim$ 120 mass region, where shell effects are small,
the apparent GDR width is expected to increase with temperature from its ground state value 
in a manner consistent with the properties of hot liquid drop  \cite{kus98}.
In contrast, the systematic experimental data show the apparent GDR width to be constant 
till  $T\sim$ 1 MeV and increases thereafter. 
The extracted apparent GDR widths at temperatures  $T<$ 1 MeV, match pretty well with the 
apparent ground state width of $^{119}$Sb (4.5 MeV, dashed line in Fig \ref{tsfm_1}) as calculated using the spreading width 
parametrization \cite{jun08} $\Gamma$ = 0.05E$^{1.6}_{GDR}$
for the small ground state deformation ($\beta$ = -0.12) \cite{mol95}.
The discrepancy between the experimental data and TSFM indicate the failure of the model 
in the present form in describing the evolution of the apparent GDR width with temperature below 1.5 MeV.
A similar suppression of the apparent width compared to TSFM was observed, in
the mass region $\sim$ 117, at still lower temperature (0.68 MeV) 
by measuring the high gamma rays from the hot fission fragments
produced in $^{252}$Cf cold fission \cite{dipu1}. 
At these low temperatures, several microscopic effects might 
play significant roles and should be incorporated properly for 
a better explanation of the experimental data. 
But, even  after incorporating  these corrections, the situation 
does not improve \cite{kus98}. It is therefore needed to have 
a re-look into the formulation of the model with the incorporation of 
any microscopic effects that may be responsible for such deviation.

We remark here that the present work establishes the fact that the apparent GDR width remains constant at 
ground state value till T $\sim$ 1 MeV and increases subsequently thereafter as the apparent 
widths are measured below and above T=1 MeV. This was not done in ref\cite{heck03} and the conclusion 
was based on assumption, as only a single apparent GDR width (with large error) at T=1 MeV was reported. 
Moreover, the data are measured simultaneously with neutron evaporation spectra to put a constraint on 
the vital level density parameter in statistical calculation. The bremsstrahlung component in the high 
energy region is also extracted experimentally instead of keeping it as free parameter. As a result, the present 
study provides a precise experimental systematics of the apparent GDR width in low T regime 
which provides an important test to the theoretical model predictions.

The microscopic phonon damping model (PDM) \cite{dang03} 
(the dotted curve in Fig \ref{tsfm_1}), though not used widely,
 better explains the trend of the data at this low temperature region. 
The model calculates the GDR width and 
the strength function directly in the laboratory frame without any need 
for an explicit inclusion of thermal fluctuation of shapes.
It has been shown that the thermal pairing effect plays an 
important role in lowering the GDR width at  $T\leq$ 2 MeV \cite{dang03}. 
However, the PDM does not have a built-in angular momentum dependence of the GDR width 
at finite temperatures and may, therefore, be used only to describe the temperature dependence at zero spin. 
Nevertheless, it is interesting to find that the PDM, which does not take into account 
the deformation of the nucleus, better explains the data compared to TSFM that emphasizes on the 
inclusion of amplitude fluctuations in shape (deformation). 
This is in complete contrast to the GDR built on the ground states as the apparent widths
are successfully described by including the nuclear ground state deformation.
Both the models describe the data and show similar behavior for $^{120}$Sn (Fig \ref{tsfm_1}) 
at higher temperatures ( $T\geq$ 1.5 MeV) but at lower temperatures show drastically 
different behavior indicating, perhaps, the presence of more dominant quantal effects.  
This exciting result opens up a new question regarding the basic understanding  of the increase 
of the apparent GDR width in the complex quantal nuclear many 
body system at low temperature and is left for further theoretical insight. 

In summary, we have presented a systematic experimental study of the apparent GDR width 
in the temperature region of 0.9 $\sim$ 1.4 MeV in $^{119}$Sb using fusion reaction with alpha particle
to explore the uncharted region of the evolution of GDR width with temperature.
The apparent GDR widths deduced from the experiment are 
inconsistent with the predictions of thermal shape fluctuation model [TSFM]. 
In fact, the GDR width appears to be constant at its ground state value till  $T\sim$ 1 MeV and 
increases thereafter whereas the TSFM predicts a gradual increase of GDR width from its ground 
state value for  $T>$ 0 MeV. The discrepancy between the experimental data and TSFM can 
be conjectured as an artifact of quantal effects of damping mechanism at low temperature.



\end{document}